\documentclass[twocolumn,english,superscriptaddress,citeautoscript,showpacs,preprintnumbers,amsmath,amssymb,prb,floatfix,footinbib]{revtex4}
\usepackage[scaled=2]{helvet}
\usepackage[T1]{fontenc}
\usepackage[latin9]{inputenc}
\usepackage{amsmath}
\usepackage{graphicx}
\usepackage{amssymb}

\makeatletter
\providecommand{\tabularnewline}{\\}

\@ifundefined{textcolor}{}
{%
 \definecolor{BLACK}{gray}{0}
 \definecolor{WHITE}{gray}{1}
 \definecolor{RED}{rgb}{1,0,0}
 \definecolor{GREEN}{rgb}{0,1,0}
 \definecolor{BLUE}{rgb}{0,0,1}
 \definecolor{CYAN}{cmyk}{1,0,0,0}
 \definecolor{MAGENTA}{cmyk}{0,1,0,0}
 \definecolor{YELLOW}{cmyk}{0,0,1,0}
 }

\usepackage[scaled=2]{helvet}\usepackage{amstext}

\makeatletter
\@ifundefined{textcolor}{}{\definecolor{BLACK}{gray}{0}\definecolor{WHITE}{gray}{1}\definecolor{RED}{rgb}{1,0,0}\definecolor{GREEN}{rgb}{0,1,0}\definecolor{BLUE}{rgb}{0,0,1}\definecolor{CYAN}{cmyk}{1,0,0,0}\definecolor{MAGENTA}{cmyk}{0,1,0,0}\definecolor{YELLOW}{cmyk}{0,0,1,0}}

\usepackage{dcolumn}
\usepackage{bm}
\usepackage{times}
\usepackage{hyperref}
\hypersetup{colorlinks=true,linkcolor=red,citecolor=blue}
\makeatother

\usepackage{babel}

\makeatother

\usepackage{babel}

\begin{document}

\title{Strong coupling of Sm and Fe magnetism in SmFeAsO as revealed by
magnetic x-ray scattering}

\author{S. Nandi}

\email{s.nandi@fz-juelich.de}

\affiliation{Jülich Centre for Neutron Science JCNS and Peter Grünberg Institut
PGI, JARA-FIT, Forschungszentrum Jülich GmbH, D-52425 Jülich, Germany}

\author{Y. Su}

\affiliation{Jülich Centre for Neutron Science JCNS-FRM II, Forschungszentrum
Jülich GmbH, Outstation at FRM II, Lichtenbergstraße 1, D-85747 Garching,
Germany}

\author{Y. Xiao}

\affiliation{Jülich Centre for Neutron Science JCNS and Peter Grünberg Institut
PGI, JARA-FIT, Forschungszentrum Jülich GmbH, D-52425 Jülich, Germany}

\author{S. Price}

\affiliation{Jülich Centre for Neutron Science JCNS and Peter Grünberg Institut
PGI, JARA-FIT, Forschungszentrum Jülich GmbH, D-52425 Jülich, Germany}

\author{X. F. Wang}

\affiliation{Hefei National Laboratory for Physical Science at Microscale and
Department of Physics, University of Science and Technology of China,
Hefei, Anhui 230026, People's Republic of China}

\author{X. H. Chen}

\affiliation{Hefei National Laboratory for Physical Science at Microscale and
Department of Physics, University of Science and Technology of China,
Hefei, Anhui 230026, People's Republic of China}

\author{J. Herrero-Martín}

\affiliation{European Synchrotron Radiation Facility, BP 220, F-38043 Grenoble
Cedex 9, France}

\author{C. Mazzoli}

\affiliation{European Synchrotron Radiation Facility, BP 220, F-38043 Grenoble
Cedex 9, France}

\author{H. C. Walker}

\affiliation{European Synchrotron Radiation Facility, BP 220, F-38043 Grenoble
Cedex 9, France}

\author{L. Paolasini}

\affiliation{European Synchrotron Radiation Facility, BP 220, F-38043 Grenoble
Cedex 9, France}

\author{S. Francoual}

\affiliation{Deutsches Elektronen-Synchrotron DESY, D-22607 Hamburg, Germany}

\author{D. K. Shukla}

\affiliation{Deutsches Elektronen-Synchrotron DESY, D-22607 Hamburg, Germany}

\author{J. Strempfer}

\affiliation{Deutsches Elektronen-Synchrotron DESY, D-22607 Hamburg, Germany}

\author{T. Chatterji}

\affiliation{Institut Laue-Langevin, BP 156, 38042 Grenoble Cedex 9, France}

\author{C. M. N. Kumar}

\affiliation{Jülich Centre for Neutron Science JCNS and Peter Grünberg Institut
PGI, JARA-FIT, Forschungszentrum Jülich GmbH, D-52425 Jülich, Germany}

\author{R. Mittal}

\affiliation{Solid State Physics Division, Bhabha Atomic Research Centre, Trombay,
Mumbai 400 085, India}

\author{H. M. Rønnow}

\affiliation{Laboratory for Quantum Magnetism, Ecole Polytechnique Fédérale de
Lausanne (EPFL), CH-1015 Lausanne, Switzerland}

\author{Ch. Rüegg}

\affiliation{Laboratory for Neutron Scattering, Paul Scherrer Institut, CH-5232
Villigen PSI, Switzerland}

\affiliation{London Centre for Nanotechnology and Department of Physics and Astronomy,
University College London, London WC1E 6BT, United Kingdom}

\author{D. F. McMorrow}

\affiliation{London Centre for Nanotechnology and Department of Physics and Astronomy,
University College London, London WC1E 6BT, United Kingdom}

\author{Th. Brückel}

\affiliation{Jülich Centre for Neutron Science JCNS and Peter Grünberg Institut
PGI, JARA-FIT, Forschungszentrum Jülich GmbH, D-52425 Jülich, Germany}

\affiliation{Jülich Centre for Neutron Science JCNS-FRM II, Forschungszentrum
Jülich GmbH, Outstation at FRM II, Lichtenbergstraße 1, D-85747 Garching,
Germany}
\begin{abstract}
The magnetic structures adopted by the Fe and Sm sublattices in SmFeAsO
have been investigated using element specific x-ray resonant and non-resonant
magnetic scattering techniques. Between 110 and 5 K, the Sm and Fe
moments are aligned along the \textit{c} and \textit{a} directions,
respectively according to the same magnetic representation $\Gamma_{5}$
and the same propagation vector (1\,0\,$\frac{1}{2}$). Below 5
K, magnetic order of both sublattices change to a different magnetic
structure and the Sm moments reorder in a magnetic unit cell equal
to the chemical unit cell. Modeling of the temperature dependence
for the Sm sublattice as well as a change in the magnetic structure
below 5 K provide a clear evidence of a surprisingly strong coupling
between the two sublattices, and indicate the need to include anisotropic
exchange interactions in models of SmFeAsO and related compounds.
\end{abstract}
\maketitle

\section{Introduction}

Following the discovery of superconductivity in LaFeAsO$_{1-x}$F$_{x}$,
with $T_{c}$\,=\,26 K,\cite{kamihara_08} an increase of the superconducting
transition temperature to above 50\,K has been achieved by replacing
La with rare-earth ($R)$ elements.\cite{chen_08,Wang_EPL_08,Ren_epl_08,Kito_08,Bos_08}
The highest transition temperature is observed in SmFeAsO$_{1-x}$F$_{x}$
($T_{c}\sim55\,$K). Interestingly, several studies on powder samples
indicate that Sm magnetic order coexists with superconductivity over
a range of fluorine doping.\cite{Torpeano_08,Ding_08,Drew_09} Muon-spin
relaxation measurements on $R$FeAsO ($R$\,=\,La, Ce, Pr, and Sm)
compounds found considerable interaction between the rare-earth and
Fe magnetism below the ordering of Fe moments ($T$$\thicksim140$\,K)
only in CeFeAsO.\cite{Maeter_09} This leads to the conclusion that
the $R$-Fe interaction may not be crucial for the observed enhanced
superconductivity in $R$FeAsO$_{1-x}$F$_{x}$. Recent neutron scattering
measurements on NdFeAsO also found an interaction between the two
magnetic sublattices, however at $T$$\thicksim15$\,K, much below
the ordering temperature of the Fe moments.\cite{Tian_10} In the
case of EuFe$_{2}$As$_{2}$,\cite{Martin_09,Xiao_09} the only known
rare-earth containing member of the $A$Fe$_{2}$As$_{2}$ ($A$\,=\,Alkaline
earth, rare-earth) family, no interaction has been found so far. Therefore,
elucidating the interaction between the two sublattices and determining
its nature is an important endeavor in understanding magnetism and
superconductivity in the $R$FeAsO family.

Due to the strong neutron absorption of Sm, the magnetic structure
determination in SmFeAsO via neutron diffraction is considerably more
challenging than of other members of the new superconductors. The
only attempt was made on a powder sample.\cite{Rayan_09} Here we
report on the first element specific x-ray resonant magnetic scattering
(XRMS) and non-resonant x-ray magnetic scattering (NRXMS) studies
of SmFeAsO to explore the details of the magnetic structure of the
parent compound and to determine the interaction between the two magnetic
sublattices. Our resonant scattering experiments show that there is
a strong interplay between Fe and Sm magnetism. Magnetic order of
Sm exists at temperatures as high as 110\,K and can be explained
by the coupling between Sm and Fe magnetism.

\section{Experimental Details}

Single crystals of SmFeAsO were grown using NaAs flux as described
earlier.\cite{Yan_09} For the scattering measurements, an as-grown
plate-like single crystal of approximate dimensions 2$\times$2$\times0.1$\,mm$^{3}$
with a surface perpendicular to the $c$ axis was selected. The XRMS
and NRXMS experiments were performed on the ID20 beamline \cite{Paolasini}
at the ESRF (European Synchrotron Radiation Facility) in Grenoble,
France at the Sm L$_{2}$, L$_{3}$ and Fe-K absorption edges and
at the Fe K-edge at beamline P09 at the PETRA III synchrotron at DESY.
The incident radiation was linearly polarized parallel to the horizontal
scattering plane ($\pi$-polarization) and perpendicular to the vertical
scattering plane ($\sigma$-polarization) for the ID20 and P09 beamlines,
respectively. The spatial cross section of the beam was 0.5\,(horizontal)$\times$0.5\,(vertical)
mm$^{2}$ for the ID20 while it was 0.2\,(horizontal)$\times$0.1\,(vertical)
mm$^{2}$ for P09. Au\,(2\,2\,0) was used at Sm L$_{2}$ edge and
Cu\,(2\,2\,0) was used for both the Sm L$_{3}$ and Fe K absorption
edges as a polarization and energy analyzer to suppress the charge
and fluorescence background relative to the magnetic scattering signal.
The sample was mounted at the end of the cold finger of a standard
orange cryostat (at ID20), a vertical field cryomagnet (at ID20) and
a displex refrigerator (at P09) with the $a$$c$ plane coincident
with the scattering plane. Measurements at ID20 were performed at
temperatures between 1.6 and 150\,K, while the lowest achievable
temperature was 5 K at P09.

\section{Experimental Results}

\subsection{Macroscopic Characterizations}

\begin{figure}
\centering{}\includegraphics[clip,width=0.5\textwidth]{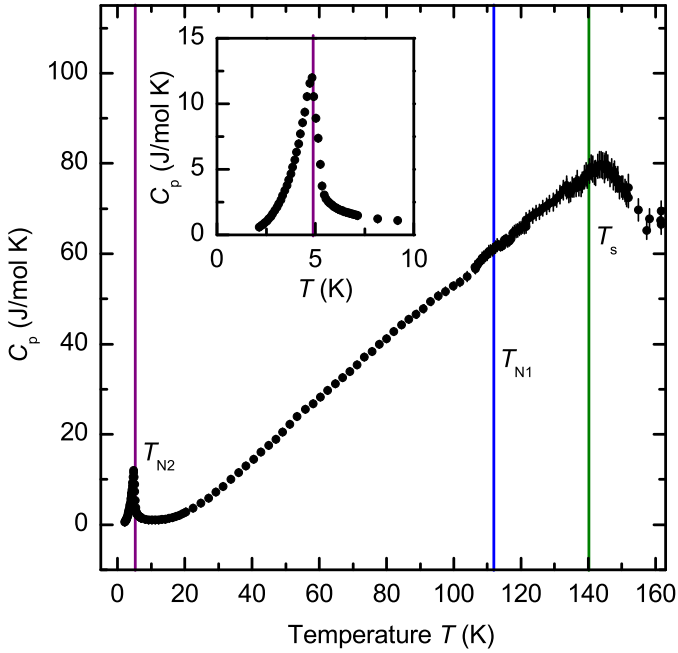}\\
 \caption{\label{fig1_specific} Temperature dependence of the specific heat.
$T_{N1}$ and $T_{N2}$ are the spontaneous magnetic ordering temperatures
of the Fe and Sm magnetic moments respectively. $T_{S}$ is the structural
phase transition temperature. Vertical lines are guides to the eye
after x-ray diffraction measurements.}

\end{figure}

Figure \ref{fig1_specific} shows the heat capacity of a SmFeAsO single
crystal, measured using a Quantum Design physical property measurement
system (PPMS). Specific heat data shows phase transitions at 143.5\,$\pm$2\,K
and 4.8\,$\pm$0.2\,K, respectively. %

\begin{figure}
\centering{}\includegraphics[clip,width=0.5\textwidth]{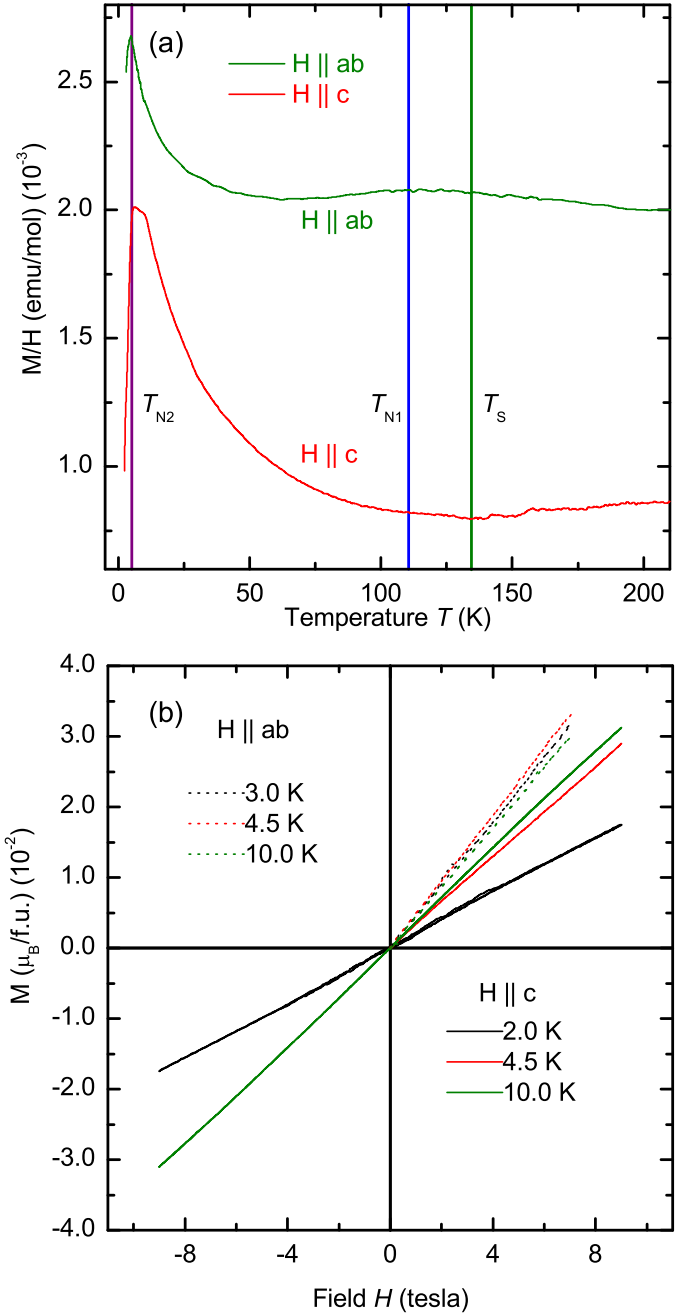}\\
 \caption{\label{magnetization}(a) Temperature dependence of the magnetic susceptibility
measured on heating of the zero-field cooled sample in a field of
1\,T. (b) \emph{M-H} curves for magnetic fields parallel and perpendicular
to the \emph{c} direction at several temperatures.}

\end{figure}

Figure \ref{magnetization}(a) shows magnetic susceptibility of a
SmFeAsO single crystal, measured using a Quantum Design SQUID magnetometer.
Magnetic susceptibility shows a clear phase transition at 5 K. There
is clear anomaly $\chi_{ab}>\chi_{c}$ over the whole temperature
range. Figure \ref{magnetization}(b) shows \emph{M-H }curves at several
temperatures for magnetic fields parallel to both \emph{c} and \emph{ab}
planes, measured using a Quantum Design vibrating sample magnetometer
(VSM). Zero field intercept of \emph{M-H} curves for both field directions
places an upper limit of ferromagnetic contribution less than 1.7$\times$$10^{-6}$$\mu_{B}$/f.u.
for all the temperatures measured.

\begin{figure}
\centering{}\includegraphics[clip,width=0.5\textwidth]{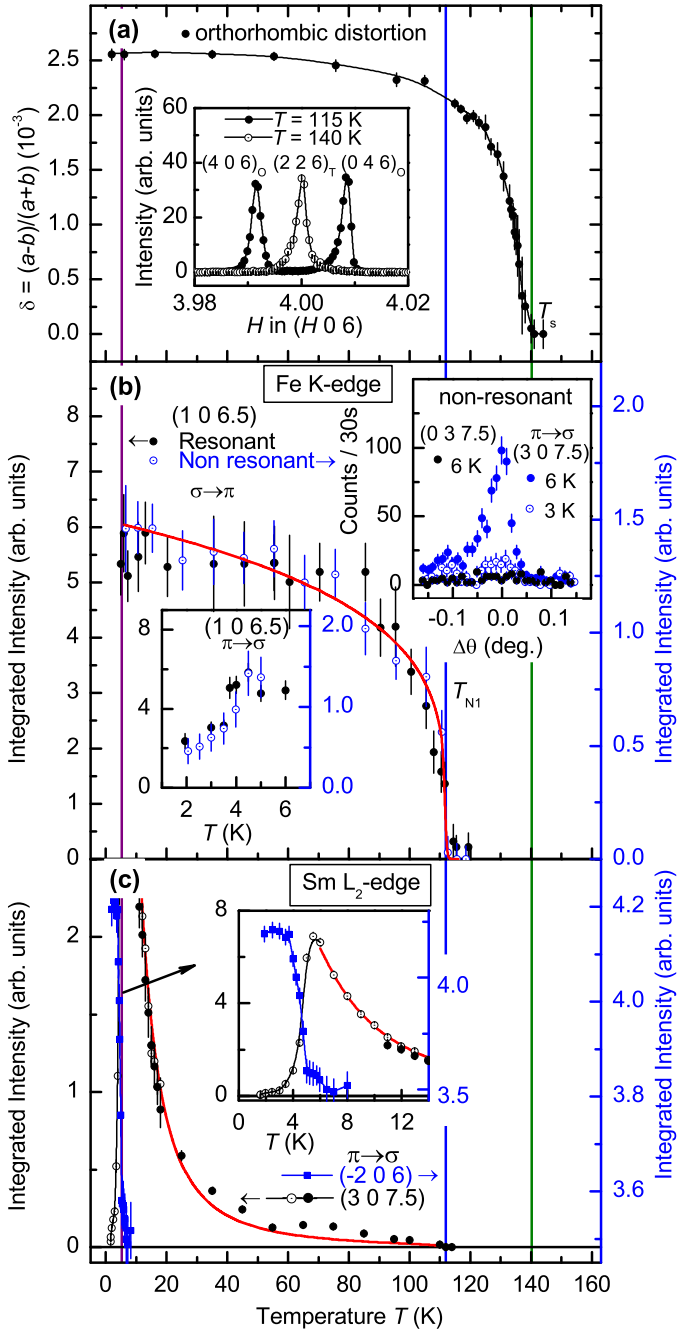}\\
 \caption{\label{fig1_tdep} (a) Temperature dependence of the orthorhombic
distortion. Inset shows ($\xi$~$0$~0) scans through the (4~0~6)
reflection. (b) Temperature dependence of the (1\,0\,6.5) reflection
measured in both resonant (at \emph{E}\,=\,7.106\,keV which is
6 eV below the Fe K-edge energy of 7.112\,keV) and non-resonant (100
eV below the Fe K-edge) conditions at P09 with a displex. Lower inset
shows temperature dependence of the (1\,0\,6.5) reflection measured
using the cryomagnet. All other measurements below 5\,K were performed
using the orange cryostat. Upper inset shows rocking scans at the
(3\,0\,7.5) and (0\,3\,7.5) reflections at selected temperatures.
(c) Temperature dependencies of the (3\,0\,7.5) and (-2\,0\,6)
reflections measured in resonant condition (\emph{E}\,=\,7.314\,keV)
at the Sm L$_{2}$ edge. Open (closed) circles represent measurements
with (without) attenuation of the primary beam. Solid thin lines serve
as guides to the eye while thick lines (red) show fit as described
in the text.}

\end{figure}

\subsection{Observation of resonant and non-resonant magnetic scattering and
characterization of the transition temperatures}

To determine whether there is a structural phase transition, as observed
in powder SmFeAsO, \cite{Martinelli_09} ($\xi\,\xi\,0)_{T}$ scans
were performed thorough the tetragonal ($T)$ (2\,2\,6)$_{T}$ Bragg
reflection as a function of temperature. As shown in the inset to
Fig.~\ref{fig1_tdep}(a), the (2\,2\,6)$_{T}$ Bragg reflection
splits into orthorhombic ($O)$ (4\,0\,6)$_{O}$ and (0\,4\,6)$_{O}$
Bragg reflections below $T_{S}$~=~140$\pm$1~K. This splitting
is consistent with the structural phase transition from space group
$P4/nmm$ to $Cmme$. The orthorhombic distortion, $\delta$, \cite{Nandi_PRL_10,Martinelli_09}
increases with decreasing temperature without any noticeable change
at the 5 K phase transition. We note that the transition temperature
$T_{S}$ is consistent with the peak observed in specific heat data.
In the remainder of the paper, we will use orthorhombic crystallographic
notation.

Below $T_{N1}$\,=\,110\,K, a magnetic signal was observed at the
reciprocal lattice points characterized by the propagation vector
(1\,0\,$\frac{1}{2}$) when the x-ray energy was tuned through the
Sm L$_{2}$ and Fe K-edges, indicating the onset of Sm and Fe magnetic
order, respectively. Figure~\ref{fig1_tdep}(b) shows a very similar
temperature evolution of the non-resonant and the resonant signal
at the Fe K-edge for the (1\,0\,6.5) reflection, supporting the
magnetic origin of the resonant signal. Resonant signal was measured
at the maximum in the resonant scattering (\emph{E}\,=\,7.106\,keV)
at the Fe K-edge while the non-resonant signal was measured approximately
100\,eV below the Fe K edge. Temperature dependence of this reflection
below 5\,K (lower inset) together with rocking scans shown in the
upper inset confirm that the iron magnetic order changes below 5 K.
Figure~\ref{fig1_tdep}(c) depicts the temperature evolution of the
(3\,0\,7.5) and (-2\,0\,6) reflections measured at the Sm L$_{2}$
edge at resonance (\emph{E}\,=7.314\,keV). At $T$$_{N2}$\,=\,5\,K,
the intensity of the (3\,0\,7.5) reflection drops quickly to zero,
and reappears at the position of the charge (-2\,0\,6) reflection,
signaling a change in the magnetic order of Sm with the magnetic unit
cell equal to the chemical unit cell. Here we note that all the measurements
below 15\,K require significant attenuation (transmission \textasciitilde{}
10\% of the incident beam) of the beam to reduce sample heating. %
\begin{figure*}
\centering{}\includegraphics[clip,width=0.95\textwidth]{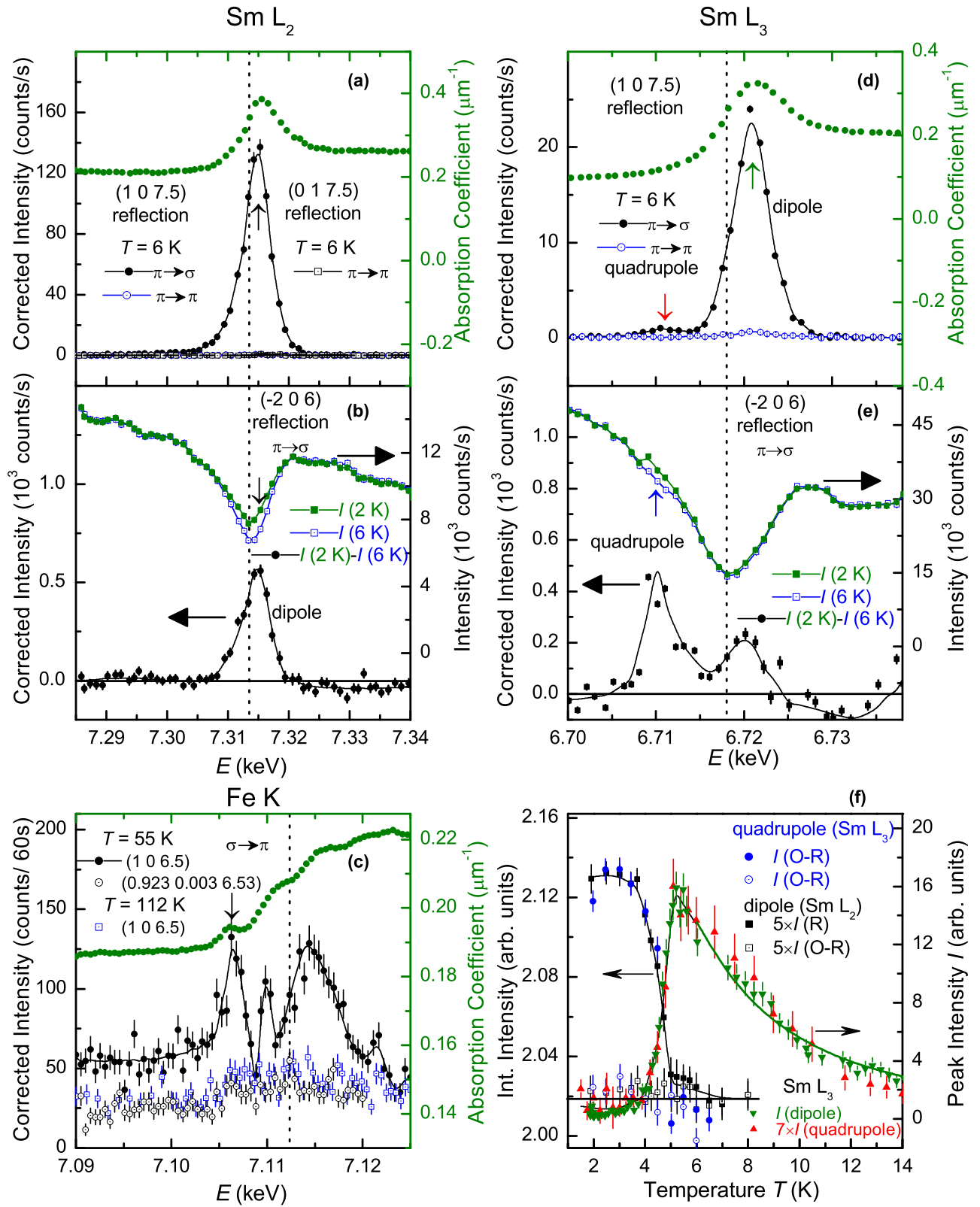}\\
 \caption{\label{EnergyScans}(a, d \& b, e) Energy scans of the (1\,0\,7.5),
(0\,1\,7.5) and (-2\,0\,6) reflections and of the absorption coefficient
at the Sm L$_{2}$ (left panel) and L$_{3}$ edges (right panel).
The dashed lines depict the Sm L$_{2}$ and L$_{3}$ absorption edges
as determined from the inflection point of the absorption coefficient.
Absorption coefficient was calculated and the intensity was corrected
following the recipe described in Refs.\,{[}\onlinecite{Sanyal_94, Luji_08}{]}.
(c) Energy scans of the absorption coefficient and of the (1\,0\,6.5)
reflection below (\textit{T\,=\,}55\,K, filled circles) and above
\textit{(T\,=\,}112\,K, open squares) $T_{N1}$ and the measured
background at \textit{T\,=\,}55\,K away from the magnetic Bragg
peak (open circles). The dashed line depicts the Fe K-edge. (f) Comparison
of the temperature dependences of the dipole and quadrupole resonances
for the (-2\,0\,6) and (1\,0\,7.5) reflections, respectively.
For the (-2\,0\,6) reflection, integrated intensity was measured
approximately 30\,eV below (off-resonance, O-R) the observed resonance
(R, \emph{E}\,=\,6.710 and 7.314\,keV for the Sm L$_{3}$ and L$_{2}$
edges, respectively) to show the temperature dependence of the pure
charge signal. The intensities have not been corrected for absorption.
In (a-e) vertical arrows indicate the energies at which temperature
dependences of the resonant signal was measured for Fig.\,\ref{fig1_tdep}(b-c)
and Fig.\,\ref{EnergyScans}(f). In (a-f) lines serve as guides to
the eye.}

\end{figure*}

To confirm the resonant magnetic behavior of the peaks, we performed
energy scans at the Sm L$_{2}$, L$_{3}$ and Fe K absorption edges
as shown in Fig.~\ref{EnergyScans}. At 6\,K, at the Sm L$_{2}$
edge we observed a dipole resonance peak approximately 2 eV above
the absorption edge for both the (1\,0\,7.5) and (-2\,0\,6) reflections.
We note that for the (-2\,0\,6) reflection charge and magnetic peak
coincide. Therefore, measurement of magnetic signal which is five
to six orders of magnitude weaker than the Thomson charge scattering
requires significant reduction of the charge background. The charge
background can be reduced significantly by a factor of $\cos^{2}$2$\theta_{analyzer}\times$$\cos^{2}$2$\theta_{sample}$
in the $\pi\rightarrow\sigma$ geometry for the reflections with the
scattering angle (2$\theta_{sample}$) close to 90$^{\circ}$.\cite{Kim_07}
(-2\,0\,6) reflection with the scattering angles (2$\theta_{sample}$)
of $\sim86^{\circ}$ and $\sim95^{\circ}$ at the Sm L$_{2}$ and
Sm L$_{3}$ edges, respectively, fulfills these conditions. The charge
signal is reduced by a factor of $\sim7\times10^{-6}$ with the scattering
angle of the analyzer (2$\theta_{analyzer}$) close to 92$^{\circ}$
for both the edges. Thus, measurement of magnetic signal seems feasible
for the (-2\,0\,6) reflection in the $\pi\rightarrow\sigma$ geometry.
Figure\,\ref{EnergyScans}(b) shows energy scans through the (-2\,0\,6)
reflection at 2 and 6\,K. Subtraction of the energy scan at 6\,K
from 2\,K shows a pronounced resonance feature at the same energy
as that observed for the charge forbidden (1\,0\,7.5) reflection.
Similar energy scans were performed at the Sm L$_{3}$ edge and are
shown in Fig.\,\ref{EnergyScans} \mbox{(d-e)}. In addition to
the dipole feature observed at the L$_{2}$ edge, quadrupole feature
appear approximately 6 eV below the Sm L$_{3}$ edge. We note that
the change in the energy spectra from the Sm L$_{2}$ to the L$_{3}$
edge is consistent with the observed resonance in another intermetallic
compound containing Sm.\cite{Adriano}

Figure.~\ref{EnergyScans}(c) shows the energy scan through the Fe
K-edge. Several features are observable in the energy spectrum: (a)
Resonant features at and above $E$\,= 7.106 keV and (b) an energy
independent non-resonant signal for energies below the resonant features.
The non-resonant signal is about a factor of 2.5 smaller than the
resonant signal. The overall energy spectrum is similar to that observed
in previous XRMS measurements in the $\sigma\rightarrow\pi$ scattering
channel at the transition metal K-edges for the BaFe$_{2}$As$_{2}$\cite{Kim_10},
Ce(Co$_{0.07}$Fe$_{0.97}$)$_{2}$ \cite{Luji_08} and NiO\cite{Neubeck}
compounds. It is noteworthy that, the pre-edge sharp resonant feature
observed at $E$\,= 7.106 keV for SmFeAsO is also present in all
of the above mentioned compounds. It appears at an energy corresponding
to the pre-edge hump observed in the respective absorption/fluorescence
spectrum. The broad resonant feature above $E$\,= 7.106 keV is also
present in all the above compounds, however, it's relative intensity
compared to the sharp feature varies from one compound to another.

Further confirmation that the dipole and quadrupole resonances at
the L$_{2}$ and L$_{3}$ edges are magnetic is obtained from the
same temperature dependence of the dipole and quadrupole resonances
as shown in Fig.~\ref{EnergyScans}(f) for both the \mbox{(-2$\,$0$\,$6)}
and (1\,0\,7.5) reflections. Since the quadrupole signal is directly
related to the ordering of the 4\emph{f} moments, the similarity of
the temperature dependences of both resonances implies that both the
dipole and quadrupole resonances are purely magnetic. %
\begin{figure}
\centering{}\includegraphics[clip,width=0.51\textwidth]{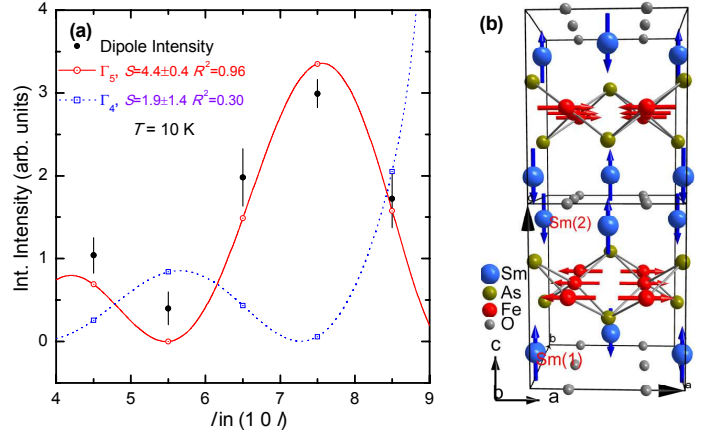}\\
 \caption{\label{TdepL3}(a) $l$ dependence of the integrated intensity at
the Sm L$_{2}$ edge along with the fits for the (1\,0\,$\frac{l}{2}$)
reflections. Open symbols are the calculated intensities. Lines serve
as guides to the eye. (b) Proposed magnetic structure in the temperature
range 5\,K $\leq T\leq110$\,K }

\end{figure}

\subsection{Magnetic structure in the temperature range $5\, K\leq T\leq110\, K$}

We now turn to the determination of the magnetic moment configuration
for the Sm moments in the temperature range $T_{N2}\leq T\leq T_{N1}$.
For the crystallographic space group $Cmme$, and propagation vector
of the form (1\,0\,$\frac{1}{2})$, six independent magnetic representations
(MRs) are possible.\cite{Wills_00} All the MRs along with the calculated
intensities for different polarization geometries are listed in Table\,
\ref{basis_vector_table_sm}. Among all the MRs, $\Gamma_{8}$\,(F)
and $\Gamma_{1}$\,(AF) MRs allow magnetic moment along $a$, $\Gamma_{2}$\,(F)
and $\Gamma_{7}$\,(AF) along $b$ and, $\Gamma_{4}$\,(F) and $\Gamma_{5}$\,(AF)
along $c$ direction, respectively. Here, F and AF denotes ferro and
antiferromagnetic alignment between Sm(1) and Sm(2) moments, respectively
(see Fig.~\ref{TdepL3}(b)). For a second-order phase transition,
Landau theory predicts that only one of the six above mentioned MRs
is realized at the phase transition.\cite{Wills_00} We note that
the $\pi\rightarrow\pi$ scattering geometry is sensitive only to
the moment perpendicular to the scattering plane for the dipole resonance.\cite{hill}
Since, no magnetic signal was observed at the (0\,1\,7.5) (sensitive
to $\Gamma_{1}$ and $\Gamma_{8}$) and (1\,0\,7.5) (sensitive to
$\Gamma_{2}$ and $\Gamma_{7}$) reflections in the $\pi\rightarrow\pi$
scattering channel at the Sm L$_{2}$ edge (see Fig.\,\ref{EnergyScans}\,(a,
d)), we can exclude the moment in the $a$ and $b$ directions and
hence, the MRs $\Gamma_{1}$, $\Gamma_{8}$, $\Gamma_{2}$ and $\Gamma_{7}$.
To differentiate between the MRs $\Gamma_{4}$ and $\Gamma_{5}$ (moment
along the $c$ direction), the integrated intensities for a series
of (1\,0\,$\frac{l}{2})$ reflections were measured (see Fig.~\ref{TdepL3}(a))
and compared with the calculated intensity as outlined below. The
intensity for a particular reflection can be written as: \begin{equation}
I=SAL|F_{m}|^{2}\label{eq:S1}\end{equation}
 where \textit{S} is arbitrary scaling factor, \textit{A}~=~$\frac{\sin(\theta+\alpha)}{\sin\theta\cos\alpha}$
is the absorption correction, \textit{L}~=~ $\frac{1}{\sin2\theta}$
is the Lorentz factor. Here, $\alpha$ is the angle that the scattering
vector \textbf{Q}\,(=$\mathbf{k_{f}}-\mathbf{k_{i}}$) makes with
the crystallographic \textbf{c} direction perpendicular to the surface
of the sample and, $\theta$ is half of the scattering angle. $\alpha$
is positive/negative for larger/smaller angles for the outgoing beam
with respect to the sample surface. $|F_{m}|$ is the modulus of the
magnetic structure factor. The magnetic structure factor $F_{m}$
for the (\emph{h}\,\emph{k}\,\emph{l}) reflections can be written
as:

\begin{equation}
F_{m}=\sum_{j}f_{j}e^{2\pi i(hx_{j}+ky_{j}+lz_{j})}\label{eq:st factor}\end{equation}
 The summation is over all the magnetic atoms in the unit cell. $f_{j}$
is the resonant/non-resonant magnetic scattering amplitude which is
listed for different polarization geometries by Hill and McMorrow
for XRMS \cite{hill} and by M. Blume and D. Gibbs for NRXMS \cite{BlumePhysRevB.37.1779}.
In particular, $f_{j}$ depends on the polarization geometry as well
as the moment direction. \emph{x}$_{j}$,\emph{ $y_{j}$} and \emph{z$_{j}$}
are the atomic position of the \emph{j}th atom within the unit cell.
The angular dependence of the magnetic structure factor originates
from the magnetic scattering amplitude $f_{j}$. For dipole resonance
and for the $\pi\rightarrow\sigma$ geometry, $f_{j}\propto~\mathbf{k_{i}}\cdot\boldsymbol{\mu}$,\cite{hill,bruckel_01}
where $\mathbf{k_{i}}$ and $\boldsymbol{\mu}$ are the wave vectors
of the incoming photons and the magnetic moment, respectively. For
the dipole resonance, and for the reflections of the type (1\,0\,$\frac{l}{2}$),
$|F_{m}|^{2}$ is proportional to $\sin^{2}\left(2\pi zl\right)\sin^{2}(\theta+\alpha)$
and $\cos^{2}\left(2\pi zl\right)\sin^{2}(\theta+\alpha)$ for the
$\Gamma_{4}$ and $\Gamma_{5}$ MRs, respectively. $z=0.137$ is atomic
position of Sm moments within the unit cell.\cite{Martinelli_09}
While $\sin^{2}\left(2\pi zl\right)$/$\cos^{2}\left(2\pi zl\right)$
term comes from the relative orientation of the magnetic moment within
the magnetic unit cell, the term $\sin^{2}(\theta+\alpha)$ comes
from the dot product between $\mathbf{k_{i}}$ and $\boldsymbol{\mu}$
{[}($90$$^{\circ}-\theta-\alpha)$ is the angle between $\mathbf{k_{i}}$
and $\boldsymbol{\mu}${]}. We note, that there is only one free parameter
for the dipole intensity (see Eqn.\,\ref{eq:S1}), namely the arbitrary
scaling factor $S$. Figure~\ref{TdepL3}(a) shows a fit to the observed
intensities for the two above mentioned MRs. Since the model calculation
with the magnetic moment in the $\Gamma_{5}$ MR closely agrees with
the observed intensity, we conclude that the magnetic Sm moments are
arranged according to the MR $\Gamma_{5}$.

For the determination of the MR for the Fe moments, the non-resonant
signal was measured at 15\,K. Similar representation analysis provides
six possible MRs for the magnetic order of Fe. All the MRs along with
the calculated intensities for different polarization geometries are
listed in Table\, \ref{basis_vector_table_fe}. Among all the MRs,
$\Gamma_{5}$ and $\Gamma_{6}$ MRs allow magnetic moment along $a$,
$\Gamma_{3}$ and $\Gamma_{4}$ along $b$ and $\Gamma_{1}$ and $\Gamma_{2}$
along $c$ direction, respectively. Among the two MRs for a particular
moment direction, the first one represents F alignment of the magnetic
moments along $b$ and AF alignment along $a$ while the second one
represents exactly the opposite alignment in the respective directions.
$\Gamma_{2}$, $\Gamma_{3}$, $\Gamma_{4}$ and $\Gamma_{6}$ MRs
can be excluded from the fact that finite intensity was observed for
the (1\,0\,6.5) reflection in the $\pi\rightarrow\sigma$ geometry
(see Table\,\ref{basis_vector_table_fe}). Zero intensities for the
(0\,3\,7.5) reflection in the $\pi\rightarrow\sigma$ geometry (see
inset of Fig.~\ref{fig1_tdep}(b)) and of the (1\,0\,6.5) reflection
in the $\pi\rightarrow\pi$ channel are also consistent with the absence
of $\Gamma_{4}$ and $\Gamma_{3}$ MRs, respectively . Finite intensity of the (1\,0\,6.5)
reflection in the $\pi\rightarrow\sigma$ channel implies that the
moments are within the $a$-$c$ scattering plane i.e. $\Gamma_{1}$
and $\Gamma_{5}$ are the possible MRs. We measured the off-specular
reflections (3\,0\,7.5) and ($\overline{3}\,0$7.5) to determine
the moment direction. The angular dependence of the non-resonant magnetic
scattering cross section for the $\pi\rightarrow\sigma$ geometry,
$f_{j}=-2\sin^{2}\theta\mathbf{k_{f}}\cdot\boldsymbol{S}$ (assuming
spin only magnetic moment of iron),\cite{BlumePhysRevB.37.1779} is
different for these two reflections providing strong sensitivity to
the moment direction. $\mathbf{k_{f}}$ and $\boldsymbol{S}$ are
the wave vectors of the outgoing photons and the spin magnetic moment,
respectively. The ratio can be written as:

\begin{equation}
\frac{I(h0\frac{l}{2})}{I(\overline{h}\,0\frac{l}{2})}=\frac{\sin(\theta-\alpha)\cos^{2}(\theta-\alpha)}{\sin(\theta+\alpha)\cos^{2}(\theta+\alpha)}\label{eq:}\end{equation}
The calculated ratio $I(3\,0\,7.5)/I(\overline{3}\,0\,7.5)$ amounts
to 5.2 and 0.35 for moments along the $a$ and $c$ directions, respectively.
The experimentally determined ratio $6.5\pm0.9$ confirms that the
moments are in the $a$ direction, i.e. the MR is $\Gamma_{5}$. We
note that this is the same MR as that of Sm, which is expected if
there is significant coupling between the two magnetic sublattices.
Arrangements of the magnetic moments according to the MR $\Gamma_{5}$
is shown in Fig.~\ref{TdepL3}(b).

\begin{table}[h]
\caption{Basis vectors for the space group $Cmme$ with ${\bf k}_{17}=(0,~1,~.5)$.
The decomposition of the magnetic representation for the $Sm$ site
$(0,~.25,~.137)$ is $\Gamma_{Mag}=1\Gamma_{1}^{1}+1\Gamma_{2}^{1}+0\Gamma_{3}^{1}+1\Gamma_{4}^{1}+1\Gamma_{5}^{1}+0\Gamma_{6}^{1}+1\Gamma_{7}^{1}+1\Gamma_{8}^{1}$.
The atoms of the nonprimitive basis are defined according to 1: $(0,~.25,~.137)$,
2: $(0,~.75,~.863)$. Lattice parameters of the orthorhombic crystal
at 100\,K \cite{Martinelli_09}: $a=5.5732$\,Å, $b=5.5611$\,Å,
$c=8.4714$\,Å.}

\begin{ruledtabular} \begin{tabular}{ccccccccc}
\multicolumn{1}{c}{IR } & Atom  & \multicolumn{3}{c}{BV components} & \multicolumn{4}{c}{Magnetic Intensity}\tabularnewline
\multicolumn{1}{c}{} &  & $m_{\|a}$  & $m_{\|b}$  & $m_{\|c}$  & \multicolumn{2}{c}{(\emph{h}\,0\,$\frac{l}{2}$)} & \multicolumn{2}{c}{(0\,\emph{k}\,$\frac{l}{2}$)}\tabularnewline
 &  &  &  &  & $\pi\rightarrow\sigma$  & $\pi\rightarrow\pi$  & $\pi\rightarrow\sigma$  & $\pi\rightarrow\pi$\tabularnewline
\hline
$\Gamma_{1}$  & 1  & 1  & 0  & 0  & Yes  & No  & No  & Yes\tabularnewline
 & 2  & -1  & 0  & 0  &  &  &  & \tabularnewline
$\Gamma_{2}$  & 1  & 0  & 1  & 0  & No  & Yes  & Yes  & No\tabularnewline
 & 2  & 0  & 1  & 0  &  &  &  & \tabularnewline
$\Gamma_{4}$  & 1  & 0  & 0  & 1  & Yes  & No  & Yes  & No\tabularnewline
 & 2  & 0  & 0  & 1  &  &  &  & \tabularnewline
$\Gamma_{5}$  & 1  & 0  & 0  & 1  & Yes  & No  & Yes  & No\tabularnewline
 & 2  & 0  & 0  & -1  &  &  &  & \tabularnewline
$\Gamma_{7}$  & 1  & 0  & 1  & 0  & No  & Yes  & Yes  & No\tabularnewline
 & 2  & 0  & -1  & 0  &  &  &  & \tabularnewline
$\Gamma_{8}$  & 1  & 1  & 0  & 0  & Yes  & No  & No  & Yes\tabularnewline
 & 2  & 1  & 0  & 0  &  &  &  & \tabularnewline
\end{tabular}\end{ruledtabular} \label{basis_vector_table_sm}
\end{table}

\begin{table}[h]
\caption{Basis vectors for the space group $Cmme$ with ${\bf k}_{17}=(0,~1,~.5)$.
The decomposition of the magnetic representation for the $Fe$ site
$(.75,~0,~.5)$ is $\Gamma_{Mag}=1\Gamma_{1}^{1}+1\Gamma_{2}^{1}+1\Gamma_{3}^{1}+1\Gamma_{4}^{1}+1\Gamma_{5}^{1}+1\Gamma_{6}^{1}+0\Gamma_{7}^{1}+0\Gamma_{8}^{1}$.
The atoms of the nonprimitive basis are defined according to 1: $(.75,~0,~.5)$,
2: $(.75,~.5,~.5)$.}

\begin{ruledtabular} \begin{tabular}{ccccccccc}
\multicolumn{1}{c}{IR } & Atom  & \multicolumn{3}{c}{BV components} & \multicolumn{4}{c}{Magnetic Intensity}\tabularnewline
\multicolumn{1}{c}{} &  & $m_{\|a}$  & $m_{\|b}$  & $m_{\|c}$  & \multicolumn{2}{c}{(\emph{h}\,0\,$\frac{l}{2}$)} & \multicolumn{2}{c}{(0\,\emph{k}\,$\frac{l}{2}$)}\tabularnewline
 &  &  &  &  & $\pi\rightarrow\sigma$  & $\pi\rightarrow\pi$  & $\pi\rightarrow\sigma$  & $\pi\rightarrow\pi$\tabularnewline
\hline
$\Gamma_{1}$  & 1  & 0  & 0  & 1  & Yes  & No  & No  & No\tabularnewline
 & 2  & 0  & 0  & 1  &  &  &  & \tabularnewline
$\Gamma_{2}$  & 1  & 0  & 0  & 1  & No  & No  & Yes  & No\tabularnewline
 & 2  & 0  & 0  & -1  &  &  &  & \tabularnewline
$\Gamma_{3}$  & 1  & 0  & 1  & 0  & No  & Yes  & No  & No\tabularnewline
 & 2  & 0  & 1  & 0  &  &  &  & \tabularnewline
$\Gamma_{4}$  & 1  & 0  & 1  & 0  & No  & No  & Yes  & No\tabularnewline
 & 2  & 0  & -1  & 0  &  &  &  & \tabularnewline
$\Gamma_{5}$  & 1  & 1  & 0  & 0  & Yes  & No  & No  & No\tabularnewline
 & 2  & 1  & 0  & 0  &  &  &  & \tabularnewline
$\Gamma_{6}$  & 1  & 1  & 0  & 0  & No  & No  & No  & Yes\tabularnewline
 & 2  & -1  & 0  & 0  &  &  &  & \tabularnewline
\end{tabular}\end{ruledtabular}

\label{basis_vector_table_fe}
\end{table}

\subsection{Temperature dependence of the magnetic intensity in the temperature
range 5\,K$\leq T\leq110.0$\,K}

Although the ordering temperatures are the same for both the Fe and
Sm sublattices, the order parameters are qualitatively different as
can be seen from Fig.\,\ref{fig1_tdep}(b-c). Particularly, the order
parameter for the Sm moment is quite unusual. Very similar temperature
dependence of the Ce sublattice magnetization in CeFeAsO has been
obtained indirectly using muon spin relaxation measurements.\cite{Maeter_09}
With reference to other systems this unusual behavior can be explained
with a ground-state doublet crystal-field level, split by an exchange
field.\cite{Sachidanandam97,Nandi_08} The Kramer's Sm$^{3+}$ions
in SmFeAsO are at the positions of local point symmetry C$_{2v}$
and, therefore, must have a doublet ground-state. At low temperatures
only the ground-state doublet is appreciably populated because the
energy difference between the ground-state and the next crystal electric-field
levels, in general, is large, and of the order of 17\,meV in case
of CeFeAsO.\cite{Chi_08} Taking into account only the ground-state
doublet and a splitting, $\triangle(T)$, we can write: \begin{equation}
m_{z}^{Sm}(T)=\frac{g_{j}\mu_{B}}{2}\tanh[\frac{\Delta(T)}{2k_{B}T}]\label{eq:msm}\end{equation}
where $\Delta(T)=g_{J}\mu_{B}B_{z}^{eff}(T)$ is the splitting of
the ground state doublet by the effective field produced by the Fe
sublattice. $g_{J}$=$\frac{2}{7}$ is the Landé g factor of the free
Sm$^{\text{3+}}$ and $\mu_{B}$ is the Bohr magneton. The effective
field should be proportional to the ordered magnetic moment of Fe,
and can be written as: \begin{equation}
B_{z}^{eff}(T)=B_{0}(1-\frac{T}{T_{N}})^{\beta}\label{eq:beff}\end{equation}
Since the observed intensity is proportional to the square of the
ordered magnetic moment ($I=Am^{2}$), $T_{N}$ (=110$\pm1$ K) and
$\beta$ (= 0.112$\pm0.008$) can be extracted from a fit to the integrated
intensity for the (1\,0\,6.5) reflection in Fig.\,\ref{fig1_tdep}(b).
A fit to the temperature dependence of the (3\,0\,7.5) reflection
in Fig.\,\ref{fig1_tdep}(c) over the whole temperature range gives
$B_{0}=(56.4\pm1.9)\:$Tesla and corresponding $\triangle(T=0)$=$(0.93\pm0.03)$
meV. We note that the value of $B_{0}$ characterizing the strength
of interaction between the two sublattices is comparable or even higher
than the value for Ce-Fe interaction in CeFeAsO.\cite{Maeter_09,Jesche_09}
The value of $\triangle$= 0.93 meV is comparable to the ground state
splitting of Ce crystal electric field levels of 0.9 meV in CeFeAsO,
measured using inelastic neutron scattering.\cite{Chi_08}

\subsection{Magnetic structure below $T\leq5.0$\,K}

For the low temperature phase ($T\leq5.0$\,K), the determination
of the magnetic structure of the Sm subsystem is considerably more
difficult due to the overlap of the magnetic intensity with the charge
intensity. Magnetization measurements with magnetic fields along the
$c$ direction and in the $a$$b$ plane exclude ferromagnetic arrangement
in the respective direction/plane, see Fig.\,\ref{magnetization}(b).
There remain three antiferromagnetic representations along $a$, $b$,
and c directions. The relative change in magnetization below 5 K is
much more pronounced for a magnetic field applied along the $c$ direction
than in the $ab$ plane (see Fig.\,\ref{magnetization}(a)). Therefore,
we conclude that the Sm moments are aligned along the $c$ direction
below 5\,K, which is in agreement with recent neutron scattering
measurements. \cite{Rayan_09}

\begin{figure}
\centering{}\includegraphics[clip,width=0.5\textwidth]{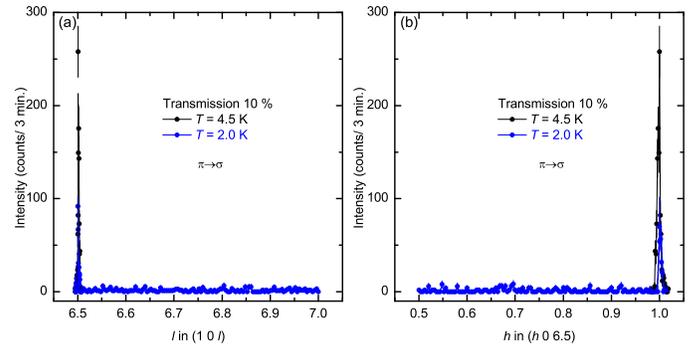}\\
 \caption{\label{H and L scans}(a \& b) \emph{l} and \emph{h} scans through
the (1\,0\,6.5) reflection. 10 \% of the incident beam was used
to reduce the sample heating. For comparison, scans at $T=4.5$\,K
is plotted together with scans taken at $T=2.0$\,K. We noticed that
the sample heating is much more in the cryomagnet than in the orange
cryostat. The remaining intensity at \emph{T}\,=\,2\,K is due to
the residual sample heating in the cryomagnet.}

\end{figure}

\begin{figure}
\centering{}\includegraphics[clip,width=0.45\textwidth]{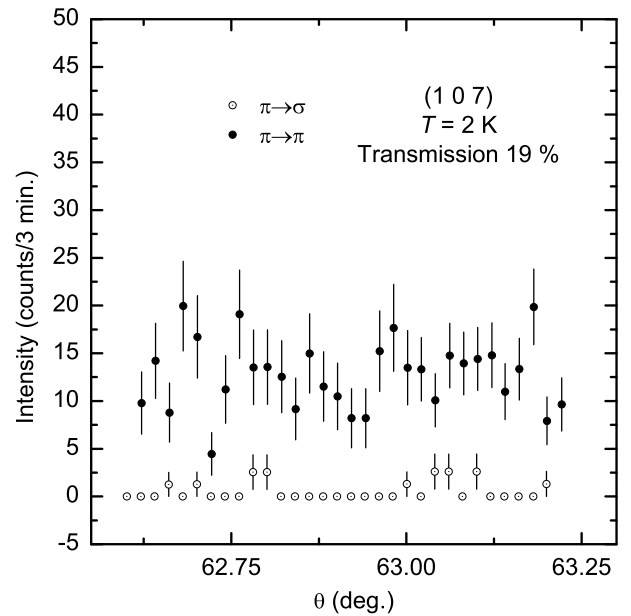}\\
 \caption{\label{rocking} Rocking scans through the (1\,0\,7) reflection
in both the $\pi\rightarrow\sigma$ and $\pi\rightarrow\pi$ channels.
Higher background in the $\pi\rightarrow\pi$ channel is mainly due
to the less suppression of the fluorescence background. }

\end{figure}
To determine the magnetic structure of the Fe moments below $T=5$\,K,
a number of possible propagation vectors suggested for the \emph{R}FeAsO
family \cite{Maeter_09}, (1\,0\,0.5), (1\,0\,0), (0.5\,0\,0.5)
and (0\,0\,0.5) were checked by rocking scans with counting time
(\textasciitilde{}3 min/data point) a factor of three larger than
other measurements at 2\,K. Measurements were performed in both the
$\pi\rightarrow\sigma$ and $\pi\rightarrow\pi$ channels to exclude
possible re-orientation of the magnetic moments from $a$ to $b$
direction. Additionally, long scans along $h$ and $l$ directions
for the (1\,0\,6.5) reflections were performed to exclude possible
incommensurate order in the respective directions, see Fig.\,\ref{H and L scans}.
However, no signal was observed for the above measurements. Its magnetic
structure with the same propagation vector as that of Sm implies a
Néel type in-plane structure which is impossible to check with hard
x-rays given the weakness of the resonant/non-resonant signal and
overlap of the magnetic signal with the charge signal. In NdFeAsO
a change in the coupling along the $c$ axis (AFM to FM) has been
observed upon the spontaneous order of Nd.\cite{Tian_10} However,
this is not the case here, as confirmed by the absence of the scattering
signal at the (1\,0\,7) reflection as shown in Fig.\,\ref{rocking}.
The absence of the scattering signal in the positions mentioned above
indicates that the in-plane as well as out-of plane correlations are
modified upon the spontaneous ordering of Sm. This observation is
unique among the $R$FeAsO family and indicate an intricate interplay
between the two sublattices. Here we note that the rare earth sites
project onto the centers of the Fe plaquettes and thus isotropic interactions
between the two vanishes by symmetry. Hence, anisotropic exchange
interactions play a major role in determining the spin structure of
the Fe sublattice and should be studied theoretically to understand
the magnetism in the $R$FeAsO family.

\section{Conclusion}

In summary, using XRMS and NRXMS we found that between 110 K and 5
K, the Sm moments are aligned in the $c$ direction while Fe moments
are aligned in the $a$ direction according to the same MR $\Gamma_{5}$
and the propagation vector (1 0 $\frac{1}{2}$). Modeling of the temperature
dependence indicates that the Sm moments are induced by the exchange
field of the Fe moments. Below 5 K, the magnetic order of both sublattices
change to a different magnetic structure, indicating an intricate
interplay between the two magnetic sublattices. Our finding of an
intricate interplay between the magnetism of Sm and Fe in the SmFeAsO
compound sheds new light on the currently debated importance of the
\emph{R}-Fe interaction in the family of iron based superconductors.

\bibliographystyle{apsrev} \bibliographystyle{apsrev}
\begin{acknowledgments}
S. N. likes to acknowledge S. Adiga, M. Angst, B. Schmitz, T. Trenit,
and S. Das for technical help.\bibliographystyle{apsrev}
\bibliography{SmFeAsO}

\end{acknowledgments}

\end{document}